\documentclass[12pt]{article}

\newcommand{\ie}{{\em i.e.}\ }
\newcommand{\eg}{{\em e.g.},\ }
\newcommand{\etal}{{\em et al.}\ }

\newcommand{\etc}{{\em etc.}\ }

\def\eps@scaling{.95}
\def\epsscale#1{\gdef\eps@scaling{#1}}
\def\plotone#1{\centering \leavevmode
\epsfxsize=\eps@scaling\columnwidth \epsfbox{#1}}

\begin{document}
\twocolumn

\section*{Digital Sky Surveys: Software Tools and Technologies}

Large digital sky surveys, over a broad range of wavelengths, both
from the ground and from space observatories, are becoming a major
source of astronomical data.  Some examples include the Sloan Digital
Sky Survey (SDSS) and the Digital Palomar Observatory Sky Survey
(DPOSS) in the visible, the Two-Micron All-Sky Survey (2MASS) in the
near-infrared, the NRAO VLA Sky Survey (NVSS) and the Faint Images of
the Radio Sky at Twenty centimeters (FIRST) in the radio.  Many others
surveys are planned or expected, in addition to the previously named
surveys.  While most surveys are exclusively imaging, large-scale
spectroscopic surveys also exist.  In addition, a number of
experiments with specific scientific goals, \eg microlensing surveys
for MACHOs, searches for near-Earth asteroids, are generating
comparable data volumes.  Typical sizes of resulting data sets (as of
the late 1990's) are in the range of tens of Terabytes of digital
information, with detections of many millions or even billions of
sources, and several tens of parameters measured for each detected
source.

This vast amount of new information presents both a great scientific
opportunity and a great technological challenge: how to process, and
calibrate the raw data; how to store, combine, and access them using
modern computing hardware and networks; and how to visualize, explore
and analyses these great data sets quickly and efficiently.  This is a
rapidly developing field, which often entails collaborative efforts
between astronomers and computer scientists.

Broadly speaking, the steps along the way include: obtaining the data
(telescope/observatory and instrument control software), processing
the data into catalogs of detected sources and their observed
properties, calibrating the data, archiving the results, and providing
the tools for their scientific analysis and exploration.

\subsection*{Survey Pipeline Software}

The actual gathering and processing of the raw survey data encompasses
many steps, which can often be performed using a software
pipeline. The first step involves hardware-specific data acquisition
software, used to operate the telescopes and the instruments
themselves.  In principle this is not very different from the general
astronomical software used for such purposes, except that the sky
surveying modes tend to require a larger data throughput, a very
stable and reliable operation over long stretches of time, and greater
data flows than is the case for most astronomical observing.  In most
cases, additional flux calibration data are taken, possibly with
separate instruments or at different times. Due to the long amounts of
time required to complete a survey (often many years), a great deal of
care must be exercised to monitor the overall performance of the
survey in order to ensure a uniform data quality.

The next step is the removal of instrumental effects, \eg flat-fielding
of CCD images, subtraction of dark current for infrared detectors.
Other than the sheer size of the data, this process is extremely
similar to the usual astronomical observational data reduction
techniques.

At this point one does some kind of automated source detection on
individual survey images, be it CCD frames, drift scans, photographic
plate scans, their subsections, or whatever other ``raw'' image format
is being generated by the survey.  This process requires a good
understanding of the noise properties, which determines some kind of a
detection significance threshold: one wants to go as deep as possible,
but not count the noise peaks.  In other words, maximize the
completeness (the fraction of real sources detected) while minimizing
the contamination (the fraction of noise peaks mistaken for real
sources).  Typically one aims for the completeness levels of at least
90\%, and contamination of less than 10\% in the first pass, and
typically the source catalogs are purified further at subsequent
processing steps.

Most source detection algorithms require a certain minimum number of
adjacent or connected pixels above some signal-to-noise thresholds for
detection.  The optimal choice of these thresholds depends on the
power spectrum of the noise.  In many cases, the detection process
involves some type of smoothing or optimal filtering, \eg with a
Gaussian whose width approximates that of an unresolved point source.
Unfortunately, this also builds in a preferred scale for source
detection, usually optimized for the unresolved sources (\eg stars)
or the barely-resolved ones (\eg faint galaxies), which are the
majority.  This is a practical solution, but with the obvious
selection biases, with the detection of sources depending not only on
their flux, but also on their shape or contrast: there is almost
always a limiting surface brightness (averaged over some specific
angular scale) in addition to the limiting flux.  The subject of
possible missing large populations of low surface brightness galaxies
has been debated extensively in the literature.  The truth is that a
surface brightness bias is always present at some level, whether it is
actually important or not.  Novel approaches to source, or more
accurately, structure detection involve so-called multi-scale
techniques.

Once individual sources are detected, a number of structural
parameters are measured for them, including fluxes in a range of
apertures, various diameters, radial moments of the light
distribution, \etc, from which a suitably defined, intensity-weighted
centroid is computed.  In most cases, the sky background intensity level 
is determined locally, \eg in a large aperture surrounding each source;
crowding and contamination by other nearby sources can present
problems and create detection and measurement biases.  Another
difficult problem is deblending or splitting of adjacent sources,
typically defined as a number of distinct, adjacent intensity peaks
connected above the detection surface brightness threshold.  A proper
approach keeps track of the hierarchy of split objects, usually called
the parent object (the blended composite), the children objects (the
first level splits), \etc  Dividing the total flux between them and
assigning other structural parameters to them are nontrivial issues,
and depend on the nature of the data and the intended scientific
applications.

Object detection and parameter measurement modules in survey
processing systems often use (or are based on) some standard
astronomical program intended for such applications, \eg FOCAS,
SExtractor, or DAOPHOT, to mention just a few of the programs often
used in 1990's.  Such programs are well documented in the literature.
Even if custom software is developed for these tasks, the technical
issues are very similar.  It is generally true that all such systems
are built with certain assumptions about the properties of sources to
be detected and measured, and optimized for a particular purpose, \eg
detection of faint galaxies, or accurate stellar photometry.  Such
data may serve most users well, but there is always a possibility that
a custom reprocessing for a given scientific purpose may be needed.

At this point (or further down the line) astrometric and flux
calibrations are applied to the data, using the measured source
positions and instrumental fluxes.  Most surveys are designed so that
improved calibrations can be reapplied at any stage.  In some cases,
it is better to apply such calibration after the object classification
(see below), as the transformations may be different for the
unresolved and the resolved sources.  Once the astrometric solutions
are applied, catalogs from adjacent or overlapping survey images can be
stitched together.

Object classification, \eg as stars or galaxies in the visible and
near-IR surveys, but more generally as resolved and unresolved
sources, is one of the key issues.  Classification of objects is an
important aspect of characterizing the astrophysical content of a
given sky survey, and for many scientific applications one wants
either stars (\ie, unresolved objects) or galaxies; consider for
example studies of the Galactic structure and studies of the
large-scale structure in the universe.  More detailed morphological
classification, \eg Hubble types of detected galaxies, may be also
performed if the data contain sufficient discriminating information to
enable it.  Given the large data volumes involved in digital sky
surveys, object classification must be automated, and in order to make
it really useful, it has to be as reliable and objective as possible,
and homogeneous over the entire survey.

In most cases, object classification is based on some quantitative
measurements of the image morphology for the detected sources.  For
example, star-galaxy separation in optical and near-IR surveys uses th
fact that all stars (and also quasars) would be unresolved point
sources, and that the observed shape of the light distribution would
be given by the point-spread function, whereas galaxies would be more
extended.  This may be quantified through various measures of the
object radial shape, \eg moments of the light distribution in
various combinations.  The problem of star-galaxy separation thus
becomes a problem of defining a boundary in some parameter space of
observed object properties, which would divide the two classes.  In
simplest approaches such a dividing line or surface is set
empirically, but more sophisticated techniques use artificial
intelligence methods, such as the Artificial Neural Nets or Decision
Trees (artificial induction software).  They require a training data
set of objects for which the classification is known accurately from
some independent observations.  Because of this additional information
input, such techniques can outperform the methods where survey data
alone are used to decide on the correct object classifications.

There are several practical problems in this task.  First, fainter
galaxies are smaller in angular extent, thus approaching stars in
their appearance.  At the fainter flux levels the measurements are
noisier, and thus the two types of objects become indistinguishable.
This sets a classification limit to most optical and near-IR surveys,
which is typically at a flux level a few times higher than the
detection limit.  Second, the shape of the point-spread function may
vary over the span of the survey, \eg due to the inevitable seeing
variations.  This may be partly overcome by defining the point-spread
function locally, and normalizing the structural parameters of objects
so that the unresolved sources are the same over the entire survey.
In other words, one must define the unresolved source template which
would be true locally, but may (and usually does) vary globally.
Furthermore, this has to be done automatically and reliably over the
entire survey data domain, which may be very heterogeneous in depth
and intrinsic resolution.  Additional problems include object
blending, saturation of signal at bright flux levels, detector
nonlinearities, \etc

The net result is that the automated object classification process is
always stochastic in nature.  Accuracies better than 90\% are usually
required, but accuracies higher than about 95\% are generally hard to
achieve, especially at faint flux levels.

In other situations, \eg where the angular resolution of the data is
poor, or where nonthermal processes are dominant generators of the
observed flux, morphology of the objects may have little meaning, and
other approaches are necessary.  Flux ratios in different bandpasses,
\ie the spectrum shape, may be useful in separating different
physical classes of objects.

\subsection*{Survey Archive Software}

Once all of the data has been extracted from the image pixels by the
survey pipeline software, it must be persisted in order to facilitate
scientific exploration. Basically, this data archiving can take one of
two forms: raw data access through the web or some suitable storage
and distribution medium (\eg CD-ROMs or data tapes), or utilization of
a database system designed specifically to handle a given digital sky
survey archive. The former storage technique, while considerably less
expensive, is certainly not optimal for any real analysis, especially
for very large data volumes. On the other hand, using a database
system provides significant advantages (\eg powerful query
expressions) limited primarily by how much a survey can afford.
Currently, most surveys are accessible from the Internet, whether it
is a simple ftp location or a full featured, web-accessible, query
engine.

While certainly flexible and generally easy to establish, web
accessible data is subject to limitations on bandwidth. Despite
technological advances which promise to ease such concerns, the rapid
growth in available data will continue to swamp available resources.
To illustrate this bandwidth problem, for a typical building network
(\eg shared Ethernet at 10 Mbit/s), it would take nearly a week and
half to sift through a Terabyte of data (which is the current
benchmark for large sky surveys), and this assumes a dedicated
bandwidth. Even at fast SCSI speeds (\eg 100 Mbit/s), it would take
approximately one day to merely move the data of interest.

Each archival center is presented with the challenging problem of
storing and serving vast amounts of complex data. Currently the
majority of the software written for these applications is in either
C++ or Java, while the actual data is transferred via ASCII, FITS, or
XML. Recently, many of the major data centers have begun to work on
sharing knowledge and expertise in order to simplify the development
process, as well as improve the overall efficacy of astronomical
archives. This work is leading to standards which dictate how archives
can communicate with each other, how archives can describe themselves
(\ie their metadata, or data that describes the data), how archives
can transfer large amounts of dynamic information, and how sources in
different archives can be cross-identified.

In general, the data processing flow is from the pixel (image) domain
to the catalog domain (detected sources with measured parameters).
This usually results in a reduction of the data volume by about an
order of magnitude (this factor varies considerably, depending on the
survey or the data set), since most pixels do not contain
statistically significant signal from resolved sources.  However, the
ability to store large amounts of digital image information on-line
opens up interesting new possibilities, whereby one may want to go
back to the pixels and remeasure fluxes or other parameters, on the
basis of the catalog information.  For example, if a source was
detected (\ie cataloged) in one bandpass, but not in another, it is
worth checking if a marginal detection is present even if it did not
make it past the statistical significance cut the first time; even the
absence of flux is sometimes useful information!

The two most common types of database management systems used within
astronomy are relational-based (where data are manipulated as tables),
and object-based (where data is manipulated individually as
objects). Each of the two methods have working sites which demonstrate
the technology in action(\eg the 2MASS project uses Informix, while
the GSCII project uses Objectivity). In general, relational systems
offer more features including third-party add-ons, and powerful query
mechanisms due to their dominant position in the business world. On
the other hand, object-based systems have shown higher performance and
better potential for scaling to extremely large data sets (\eg CERN is
developing an object-based persistence solution to multi-Petabyte
archives).

Regardless of how the data is actually persisted, with the advent of
the Internet, essentially all archives are Web accessible. This trend
away from tabular or media based archiving produces datasets which are
distributed in nature, allowing users and tools equal access to vast
amounts of data, which in the past were nearly impossible to
efficiently query. As a result, astronomical data is now able to be
utilized in a more democratic fashion resulting in uses which the
survey institutions did not even imagine.

\subsection*{Scientific Analysis and Exploration}

The end goal of digital sky surveys is to enable scientific investigations.
In some cases, the goals are so specific that the best approach is to use
analysis software designed and optimized to tackle them, \eg as in the
studies of the cosmic microwave background, in gravitational microlensing
experiments, \etc  However, most of the modern survey data sets are so
information-rich, that a wide variety of different scientific studies can
be done with the same data.  This entails some general tools for the
exploration, visualization, and analysis of large survey data sets.

The tools which will be utilized with the new datasets will need to
employ cutting-edge techniques from computer science. For example,
clustering techniques to detect rare, anomalous, or somehow unusual
objects, \eg as outliers in the parameter space, to be selected for
further investigation (\ie follow-up spectroscopy). Other examples
include genetic algorithms to improve current detection and supervised
classification methods, new data visualization and presentation
techniques, which can convey most of the multidimensional information
in a way more easily grasped by a human user, and the use of
semi-autonomous AI or software agents to explore the large data
parameter spaces and report on the occurrences of unusual instances or
classes of objects.

\def\plotfiddle#1#2#3#4#5#6#7{\centering \leavevmode
\vbox to#2{\rule{0pt}{#2}}
\includegraphics{#1}}

\begin{figure*}
\plotfiddle{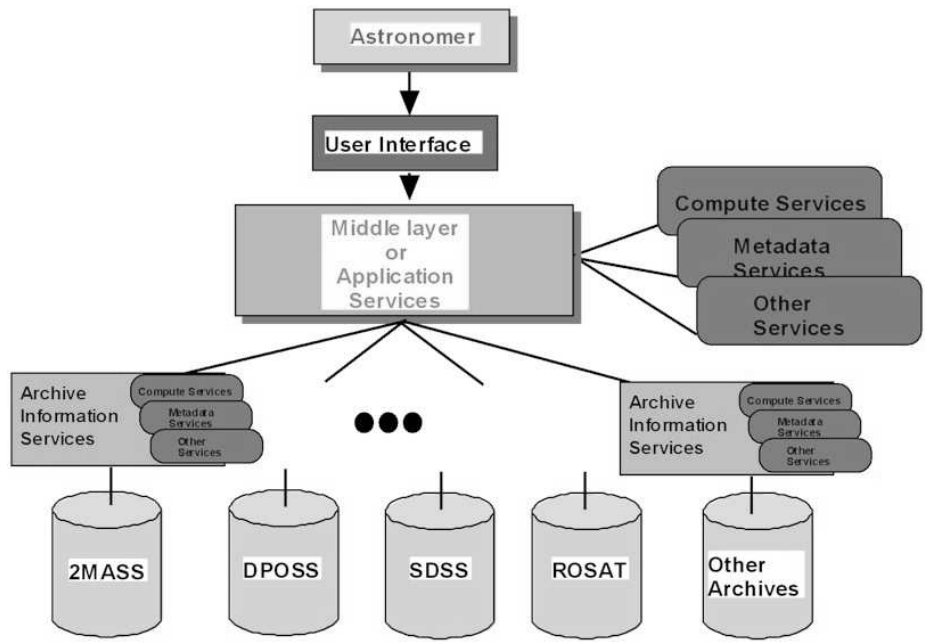}{216pt}{0}{65}{65}{-198}{-144}
\caption{
A sample prototype blueprint for the system architecture of a virtual
observatory, connecting the scientific users with a multitude of separate
digital sky survey archives (schematically represented at the bottom of
the picture).  
The basic task of such a system is to combine large volumes of data from
different archives, e.g., through the positional matching of detected 
sources.  Once this is achieved, various other computing services provided 
by such a virtual observatory may include visualisation of data, data
mining, clustering analyis, etc.
This system model is predicated on the universal adoption of standards
dictating everything from how archives communicate with each other to how 
data are transferred between archives, services and users. 
\label{as}}
\end{figure*}

The nature of sky surveys and the resulting data sets dictates the
kinds of science one may wish to do with them.  For example, most
surveys would cover a large solid angle, but not go as deep as the
typical ``pointed observations''.  This, in turn, drives the kinds of
data analysis software needed for their scientific exploration.
Broadly speaking, the kinds of astronomical investigations for which
surveys are well suited include the following:

\begin{description}

\item[\mbox{Multiwavelength Astronomy \hspace*{\columnwidth}}]

Combining surveys done at different wavelengths, for a more
panchromatic view of the universe.  Typically, optical IDs are needed
for most of the follow-up work.  An obvious example are the optical
identifications of radio or x-ray sources.  A typical basic software
requirement is efficient and accurate matching of sources detected in
different wavelengths.  If positional accuracies and source densities
in matched surveys are comparable this is a relatively straightforward
task; however, if either of these two conditions is not satisfied,
more sophisticated algorithms may be needed to assign matching
probabilities to multiple candidates. These techniques can generally
be optimized for certain classes of sources by utilizing as much a
priori knowledge as possible.

\item[\mbox{Statistical Astronomy \hspace*{\columnwidth}}]

Studies of the large-scale structure (if one detects galaxies in
copious numbers), or studies of the Galactic structure (if one detects
stars in copious numbers).  The sheer numbers of detected sources make
the Poissonian fluctuations unimportant, but systematic errors may
limit the results.  Accurate and uniform flux calibrations and source
classifications are essential for such applications, and possible
biases should be modeled. In addition, the sheer size of most surveys
allows for further subdivision of various analysis (\eg by
morphological or spectral type) for more specific results.

\item[\mbox{Rare Object Searches \hspace*{\columnwidth}}]

Searches for rare types of objects or groups of objects, whose
frequency of occurrence is so low that huge input source catalogs are
necessary in order to find them. This ``needle in a haystack'' search
is facilitated by the vast number of sources in current surveys (\ie a
one in a million object would appear a thousand times in a billion
source survey).  Typically such rare objects are found as outliers in
some parameter space, e.g., colors: the bulk of ``ordinary'' sources
(normal stars, galaxies, \etc) would form well-defined clusters in the
parameter space of observables, and some rare or peculiar types of
objects may be found away from them.  Unsupervised classification and
cluster analysis techniques may be especially useful for this task.
This data-mining also facilitates significantly higher efficiencies in
performing follow-up spectroscopy of unusual objects.  We note that
rare objects may belong to known classes (\eg distant quasars, brown
dwarfs, etc.), which can be known to occupy some {\it a priori}
determined portion of the given parameter space.  Alternatively, there
may be previously unknown or unexpected types of objects discovered in
this manner.

\end{description}

\subsection*{Future Directions}

Astronomy finds itself at an interesting time, as several, large area,
digital sky surveys are currently underway. In the future, the number
of such surveys, as well as the wavelength coverage, will only
continue to increase, providing temporal as well as spatial coverage
of the sky. This flood of data necessitates a new approach to data
handling -- a virtual observatory (see Figure 1). Individual surveys
are important in their own right, but the federation of multiple,
cross-wavelength digital surveys provides a tremendous opportunity to
truly quantify the origins of stars, galaxies and the universe itself.

This future situation provides the opportunity to adopt a new research
paradigm for studying the heavens, as the ability to perform cutting
edge research will not be restricted to those fortunate enough to have
access to the best facilities. Instead, anyone who has the diligence
and ability to sift through the avalanche of data can perform novel
science.

\subsection*{Further Readings}

The field of large digital sky surveys, the related software methodologies, 
and the new astronomy they enable are developing very rapidly as of this 
writing (late 1999).  There are thus no standard texts or reviews, and 
whatever articles do exist tend to become obsolete very quickly.  Probably 
the best way to find more up to date information is through the World Wide Web 
(and the Interested Reader will know how).  While many web sites can be
somewhat ephemeral, we suggest one which will provide a number of useful links
in this field, and which we hope will be active throughout the useful shelf
life of this Encyclopedia: {\tt http://www.digital-sky.org} 

Another possibility is conference proceedings on this and related subjects,
\eg the series of volumes on ``Astronomical Data and Software Systems'',
published by the Astronomical Society of the Pacific in their Conference
Series.  A number of useful papers can also be found in the proceedings of 
IAU Symposium 179, ``New Horizons From Multi-Wavelength Sky Surveys'',
eds. B.J.~McLean \etal, Dordrecht: Kluwer Academic Publ. (1998).

An example of a particular software system for processing of digital sky
surveys is described by Weir, N., \etal 1995, 
{\it Publ. Astron. Soc. Pacific} {\bf 107}, 1243.

\begin{flushright}
\large{
S. George Djorgovski\\
Robert J. Brunner\\
}
\end{flushright}
\end{document}